# $Fe^{2+}$:$Zn_{1-x}Mg_xSe$ single crystals, a new material for active elements of tunable lasers for 4-5 µm

Yu. A. Zagoruiko, N.O. Kovalenko, O.A. Fedorenko, A.S. Gerasimenko

*Institute for Single Crystals, STC "Institute for Single Crystals", National Academy of Sciences of Ukraine, 60 Lenin Ave., 61001 Kharkiv, Ukraine*

## Abstract

Proposed is a new laser material for the making of active elements of tunable lasers for mid IR region (4-5 µm) – $Zn_{1-x}Mg_xSe$ ($0.11<x<0.42$) single crystals doped with $Fe^{2+}$ ions. The optical transmission spectra of $Fe^{2+}$:$Zn_{0.89}Mg_{0.11}Se$ and $Fe^{2+}$:$Zn_{0.69}Mg_{0.31}Se$ samples contain strong absorption bands with maxima at 3.212 and 3.374 µm, respectively. The generation band maximum of the wide-gap semiconductor laser material $Fe^{2+}$ :$Zn_{1-x}Mg_xSe$ shifts towards longer wavelengths with the rise of the concentration of Mg in the crystalline matrix.

***Key words:*** $Fe^{2+}$:$Zn_{1-x}Mg_xSe$, active laser element, generation band, medium IR region

Nowadays much attention is being paid to the creation of a new class of laser crystals meant for the making of active elements of tunable lasers for mid infrared region (2…5 µm). Such materials can be obtained e.g. by doping of binary $A^{II}B^{VI}$ compounds and their solid solutions with ions of transition metals ($Cr^{2+}$, $Fe^{2+}$, $Co^{2+}$, etc.) [1,2]. Among the said crystals, $Cr^{2+}$: ZnSe and $Fe^{2+}$: ZnSe are studied most thoroughly [1,2,3]. In particular, $Fe^{2+}$: ZnSe crystals containing $Fe^{2+}$ ions with $1\cdot10^{18}$ $cm^{-3}$ concentration were found to be a promising laser medium for 4-5 µm wavelength range. Smooth change of the generation wavelength of $Fe^{2+}$: ZnSe lasers in 3.77-4.40 µm spectral region using dispersion prismatic resonator was reported in [2].

We were the first to show the possibility to use chromium-doped $Zn_{1-x}Mg_xSe$ single crystals (possessing higher thermal stability and wider energy gap in comparison with ZnSe) for the obtaining of a material for active elements of tunable lasers for mid infrared [4]. The generation characteristics of such lasers were studied in [5] using $Cr^{2+}$: $Zn_{0.76}Mg_{0.24}Se$ as an example. The maximum of the generation band in the new thermally stable wide-band gap hexagonal material $Cr^{2+}:Zn_{1-x}Mg_xSe$ was found to correspond to 2.47 μm wavelength, thus exceeding the said parameter of all the known active media based on $Cr^{2+}$ doped binary $A^{II}B^{VI}$ compounds and their solid solutions: ZnS; ZnSe; ZnTe; CdS; CdSe; CdTe; $Cd_{0.9}Zn_{0.1}Te$; $Cd_{0.65}Mg_{0.35}Te$; $Cd_{0.85}Mn_{0.15}Te$; $Cd_{0.55}Mn_{0.45}Te$.

To widen the nomenclature of the materials meant for the production of active elements of tunable lasers working in 4…5 μm wavelength range, and to continue the investigations reported in our papers [4,5], we propose to use $Fe^{2+}:Zn_{1-x}Mg_xSe$ ($0.11<x<0.42$) single crystals as a new active laser medium.

The study was carried out on $Fe^{2+}:Zn_{1-x}Mg_xSe$ crystals 23 mm in diameter and 50 mm high which contained $2\cdot10^{-2}$ wt.% of iron. The crystals were grown by the vertical Bridgman method in graphite crucibles under excessive pressure of argon. The starting material consisted of polycrystalline ZnSe, MgSe and FeSe compounds.

Shown in Fig.1 is the external appearance of the investigated sample.

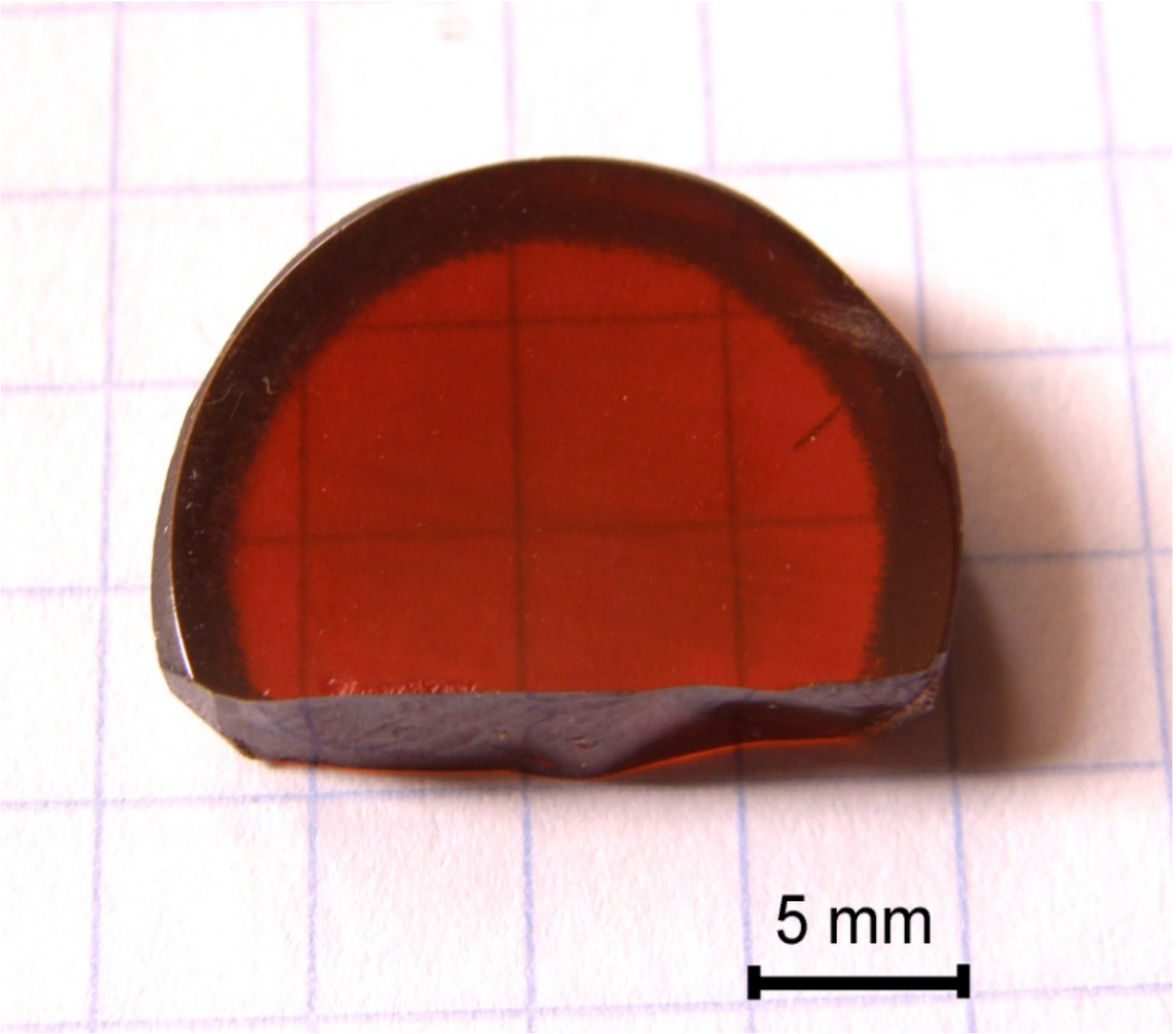

Fig.1. External appearance of the investigated sample.

The optical absorption spectra of crystalline $Fe^{2+}$: ZnSe and $Fe^{2+}$: $Zn_{1-x}Mg_xSe$ are presented in Fig. 2.

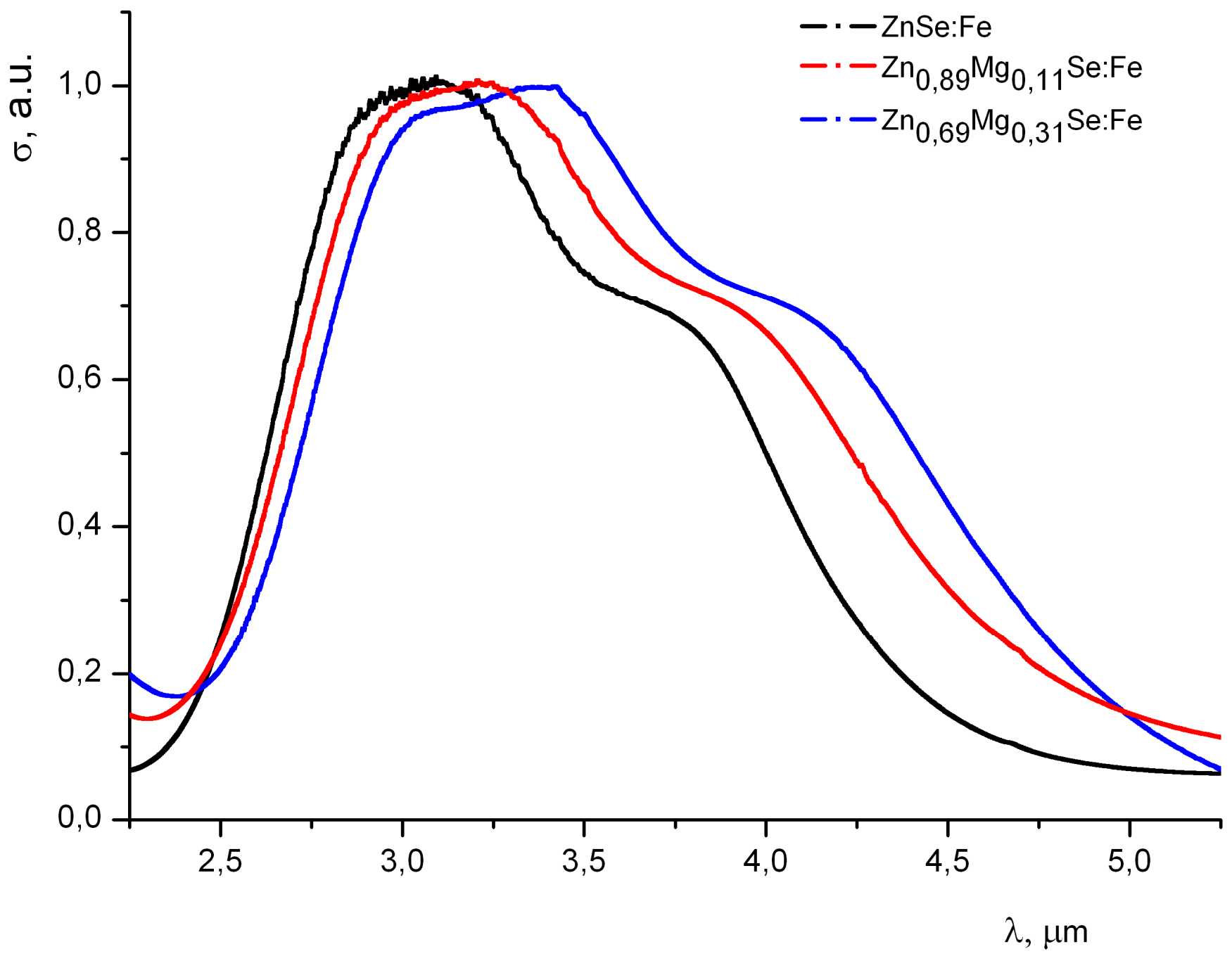

Fig.2. Optical absorption spectra of $Fe^{2+}$:$Zn_{0.89}Mg_{0.11}Se$ and $Fe^{2+}$:$Zn_{0.69}Mg_{0.31}Se$ samples.

These spectra were recalculated, taking into account the corresponding refraction coefficients, from the optical transmission spectra of the crystals under consideration. The transmission spectra were measured by a Perkin Elmer Spectrum One FT-IR spectrometer. For comparison, the data for $Fe^{2+}$ :ZnSe are also shown in this figure.

As seen from Fig. 2, the optical transmission spectra of $Fe^{2+}$:$Zn_{0.89}Mg_{0.11}Se$ and $Fe^{2+}$:$Zn_{0.69}Mg_{0.31}Se$ samples contain strong absorption bands with maxima at 3.212 and 3.374 µm, respectively, caused by the presence of $Fe^{2+}$ ions in the crystalline matrix. For $Fe^{2+}$:ZnSe the absorption band maximum is observed at 3.106 µm. Such absorption bands testify to the possibility to use $Fe^{2+}$:$Zn_{1-x}Mg_xSe$ single crystals as a new thermally stable material for active elements of tunable lasers for mid IR region. The generation band of the crystals is shifted towards longer wavelengths with respect to the one of $Fe^{2+}$:ZnSe. It is also seen that the location of the absorption band maximum of the wide-gap semiconductor laser material $Fe^{2+}$:$Zn_{1-x}Mg_xSe$ depends on the concentration of magnesium in the matrix:

with the rise of this concentration the absorption band maximum shifts to the long-wavelength region.

Thus, the obtained results are of practical interest, since they point to the possibility to shift the generation band of tunable $Fe^{2+}:Zn_{1-x}Mg_xSe$ lasers towards longer wavelengths approximately by 0.27 μm (Fig. 2), and to widen the spectral range of the change of their generation wavelength, as it has been observed for $Cr^{2+}:Zn_{0.76}Mg_{0.24}Se$ [4,5,6].